\documentclass[twocolumn,sort&compress,5p]{elsarticle}
\usepackage{geometry}
\usepackage{amssymb}
\usepackage{amsmath}
\usepackage{epsfig}
\usepackage{tabularx}
\usepackage{inputenc}
\usepackage{url}
\usepackage{breqn}
\usepackage{hyperref}
\usepackage{footmisc}
\usepackage{titlesec}

\makeatletter
\def\ps@pprintTitle{MI-TH-1940%
   \let\@oddhead\@empty
   \let\@evenhead\@empty
   \def\@oddfoot{\reset@font\hfil\thepage\hfil}
   \let\@evenfoot\@oddfoot
}
\makeatother

\usepackage{graphicx}
\usepackage{caption}
\usepackage{slashed}
\usepackage{wrapfig}
\usepackage{color}
\usepackage{fancyhdr}

\title{A heavy neutral gauge boson near the $Z$ boson mass pole via third generation fermions at the LHC}

\author{Mohammad Abdullah}
\author{Mykhailo Dalchenko}
\author{Teruki Kamon}
\author{Denis Rathjens}
\author{Adrian Thompson\corref{cor1}}

\cortext[cor1]{Corresponding author}

\address{Mitchell Institute for Fundamental Physics and Astronomy, Department of Physics  and Astronomy, \\ Texas A$\&$M University,
College Station, TX 77843}
\begin{document}

\begin{abstract}
We explore the physics of a new neutral gauge boson, ($Z^\prime$), coupling to only third-generation particles and with a mass near the electroweak gauge boson mass poles. A $Z^\prime$ boson produced by top quarks and decaying to tau leptons is considered. With a simple search strategy inspired by existing analyses of the standard model gauge boson production in association with top quarks, we show that the Large Hadron Collider has good exclusionary power over the model parameter space of the $Z^\prime$ boson even at the advent of the high-luminosity era. It is shown that the $t\bar{t}Z^\prime$ process allows one to place limits on right-handed top couplings with a $Z^\prime$ boson that preferentially couples to third generation fermions, which are at present very weakly constrained.
\end{abstract}

\maketitle

\section{Introduction}
\label{sec:intro}
A new neutral massive vector gauge boson, designated as $Z'$, is considered to be a relatively low-hanging fruit for resonance searches at the CERN Large Hadron Collider (LHC). Due to the numerous allowed combinations of couplings to standard model (SM) fermions, however, there remain production and decay modes that are yet to be thoroughly explored. Of particular interest, in light of the flavor hierarchy in the SM, is a $Z'$ that couples preferentially to the third generation fermions \cite{Holdom:1994ss, Frampton:1996bs, Muller:1996dj, Andrianov:1998hx, Babu:2017olk, Kamada:2018kmi, Elahi:2019}.

So far, the experimental coverage of third generation couplings in flavor non-universal $Z^\prime$ and their associated resonances has been mainly limited to heavy masses; $m_{Z^\prime} \gtrsim m_t$ \cite{Aad:2015osa, CMS:2016zxk, Sirunyan:2017uhk, Haller:2017hjf, ATLAS:2017mpg}. Recently, searches for light $b\bar{b}$ \cite{Sirunyan:2018ikr} resonances and $\tau^- \tau^+$ \cite{CMS:2019hvr} resonances in association with $b$-jets have been performed, setting limits for the first time on these signatures. In addition to the inherent difficulties associated with third generation final states, resonances in the low mass regime are more challenging, especially in the vicinity of the electroweak gauge boson mass poles. Since the present limits are modest and ready to be improved over the upcoming run cycle of the High Luminosity (HL)-LHC, the following question is open: is there any new and exclusively third generation physics around the scale of the electroweak gauge boson masses?

To this end, and in contrast with existing studies, we propose a search for a $Z^\prime$ that is produced through top quarks and decays primarily to tau leptons. The negligibly low presence of top quarks in the proton's parton distribution functions (PDFs) at the LHC energy range leads us to consider $Z^\prime$ production either through gluon splitting followed by top fusion, or top production followed by $Z^\prime$ emission. In both production mechanisms, the resonant ditau is supplemented with a top quark pair which, convolved with reliable top tagging, provides an additional handle in discriminating the signal from the background. Relevant studies can be found in Refs. \cite{Sirunyan:2017uhk, Hill:1991at, Hill:1993hs,Hill:1994hp, Harris:2011ez,Rosner:1996eb, Lynch:2000md, Carena:2004xs, Choudhury:2007ux, Khachatryan:2015sma, CMS:2018ohu, Aaboud:2018mjh, Sirunyan:2017yar, Cerrito:2016qig, Arina:2016cqj, Pedersen:2015knf, Fox:2018ldq, Alvarez:2019uxp, Boucenna:2016qad}.

For the pragmatically oriented reader, this search is interesting in its own right in how it fits into the landscape of resonance searches at colliders. From a model building stand point, however, a model with only such interactions and particle content is not anomaly free and needs to be supplemented with additional physics. There are numerous ways to do so. For example in Ref.~\cite{Fox:2018ldq}, where a heavy spectator fermion partner to the top quark is responsible for anomaly cancellation. Within such models, $\mathcal{O}(1)$ values of the $Z^\prime$ couplings to the SM fermions are attainable. Each model has the potential of altering our limits and introducing additional constraints in different ways. Rather than restricting ourselves to one such specific model, we will opt for a minimal setup where the $Z'$ couples only to right handed top quarks and right handed tau leptons, since this coupling scheme is not strongly constrained by existing experiments. We will provide the reader with the mathematical formalism needed to translate our results to a wider class of models.

This paper is organized as follows: in Section \,\ref{sec:model} we present our simplified model, in Section \,\ref{sec:search} we detail our search strategy, in Section \,\ref{sec:results} we present our findings, and in Section \,\ref{sec:conclusion} we conclude.

\section{The Model}
\label{sec:model}
We consider a toy model that extends the SM Lagrangian with an additional $U(1)^\prime$ gauge symmetry in the third generation of the SM fermions and introduces a new gauge vector boson $Z^\prime$. The new physics couplings are given as follows:
\begin{eqnarray}
\label{eq:lag}
\mathcal{L}&\supset& Z^{ \prime \mu} \, \bigg[  \,g_{\tau} \bar{\tau} \,\gamma_{\mu} (c_{\tau L} P_L + c_{\tau R} P_R) \,\tau+\,g_{\nu} \bar{\nu_\tau} \,\gamma_{\mu} (c_{\nu L} P_L) \,\nu_\tau \nonumber \\
&+& \,g_{t}  \bar{t} \,\gamma_{\mu} (c_{t L} P_L + c_{t R} P_R) \,t+\,g_{b}  \bar{b} \,\gamma_{\mu} (c_{b L} P_L+ c_{b R} P_R) \,b \nonumber \bigg]\, \\ 
&+& \, \textrm{h.c.},
\end{eqnarray}
where $g_\tau$, $g_\nu$, $g_t$ and $g_b$ are new physics couplings, and $c_{fX}$ ($X=L,\;R$) are the strengths of left handed and right handed couplings for fermion $f$. For the case we are interested in we set $c_{tR} = c_{\tau R} = 1$ and all other coefficients $c_{fX}$ to zero.  For the purpose of rescaling the signal, $\sigma(pp \rightarrow t\bar{t} Z') \times BR(Z' \rightarrow \tau^+ \tau^- )$, it is useful to know the decay width of $Z'$ which, in the massless fermion approximation, is given by:
\begin{equation}
\Gamma (Z' \rightarrow f\bar{f}) = \frac{N_c}{24 \pi } g_f^2  {m_{Z'}} \left( {c_{fL}}^2+ {c_{fR}}^2\right),
\end{equation}
where $m_{Z'}$ is the $Z'$ mass, $N_c = 1$ for leptons, and and $N_c = 3$ for quarks. Note that, as written, the expression is valid for general values of $c_L$ and $c_R$, not just for 0 and 1.

If a bottom quark coupling is included, then the dominant constraint for a light $Z^{\prime}$ boson comes from the added partial width to the bottomonium decay $\Upsilon (1S) \to \tau \tau$. The latest central value of 
\begin{equation}
R_{\tau\mu} = \dfrac{\Gamma_{\Upsilon(1S)\to\tau\tau}}{\Gamma_{\Upsilon(1S)\to\mu\mu}} = 1.005 \pm 0.013 (\text{stat}) \pm 0.022 (\text{syst}) \nonumber
\end{equation}
was reported by the BaBar Collaboration \cite{PhysRevLett.104.191801}. The SM prediction including one-loop QED diagrams and lepton mass effects is $R^{SM}_{\tau\mu} = 0.9924 \pm O(10^{-5})$ \cite{Aloni:2017eny}. For the couplings in Equation~\ref{eq:lag}, $R_{\tau\mu}$ is modified to
\begin{align}
\label{eq:rtaumu}
R_{\tau\mu}= R_{\tau\mu}^{SM}\bigg[1&+ (c_{b L}+c_{bR})(c_{\tau L}+c_{\tau R})\dfrac{3 g_q g_\tau m_\Upsilon^2}{8 \pi  \alpha (m_{Z^\prime}^2 - m_\Upsilon^2)}\nonumber \\
& + \mathcal{O}\bigg(\dfrac{g_q^2 g_\tau^2 m_\Upsilon^4}{m_{Z^\prime}^4}\bigg)\bigg]
\end{align}
where $\alpha$ is the fine-structure constant and $m_\Upsilon \approx 9.46$ GeV. If the $Z^\prime$ contributes maximally to the upper-1$\sigma$ $R_{\tau\mu}$ value from the BaBar experiment (also including the second-order contribution), Equation~\ref{eq:rtaumu} translates to the following bound:
\begin{equation}
\label{eq:constraint}
(c_{b L}+c_{bR})(c_{\tau L}+c_{\tau R})\bigg(\dfrac{g_q g_\tau}{0.01}\bigg) \bigg(\dfrac{40 \, \text{GeV}}{m_{Z^\prime}}\bigg)^2 \lesssim 3.29 
\end{equation}

Note that if the $Z^\prime$ couples only to the right-handed projections of the third generation quarks, then the top and bottom couplings may be varied independently of each other in certain ultraviolet completions. The $\Upsilon(1S)$ decay bounds are presented here assuming the couplings are equal.

We now proceed to detail our search strategy for the $t\bar{t}Z^\prime(\tau\tau)$ final state. The complementary search for $b\bar{b}Z^\prime(\tau\tau)$ had been thoroughly investigated in the context of a $(B-L)_3$ model \cite{Elahi:2019}.

\section{Search Strategy}
\label{sec:search}
For a quark-produced $Z^\prime$ decaying to tau leptons, QCD and $t\bar{t}$ events are the first SM backgrounds of consideration; they can easily dwarf the signal process with their $\gtrsim 1000$ pb cross-section at $\sqrt{s} = 13$ TeV. If hadronically-decaying tau leptons $\tau_h$ are within the event selection, the $\tau_h$ fake rate for jets of about $0.5-1$ \% ~\cite{Sirunyan:2018pgf} has substantial effect for such large cross-sections. However, the presence of light leptons ($\ell = e$, $\mu$) with $p_T\gtrsim 25$ GeV in the event selection can help mitigate those backgrounds.

\begin{figure}[!tbh]
  \begin{center}
   \includegraphics[width=0.23\textwidth]{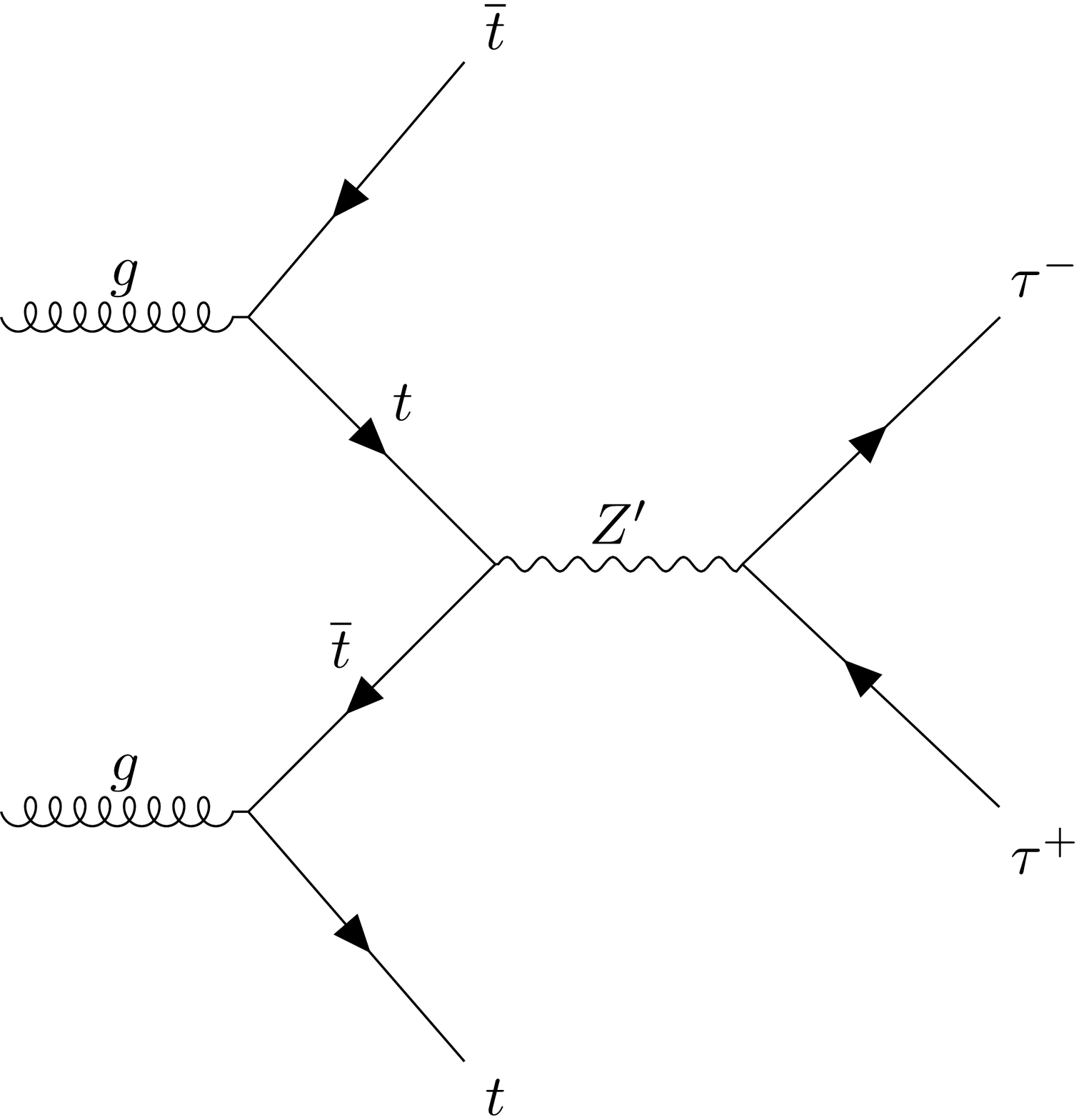}
\raisebox{0.35\height}{\includegraphics[width=0.23\textwidth]{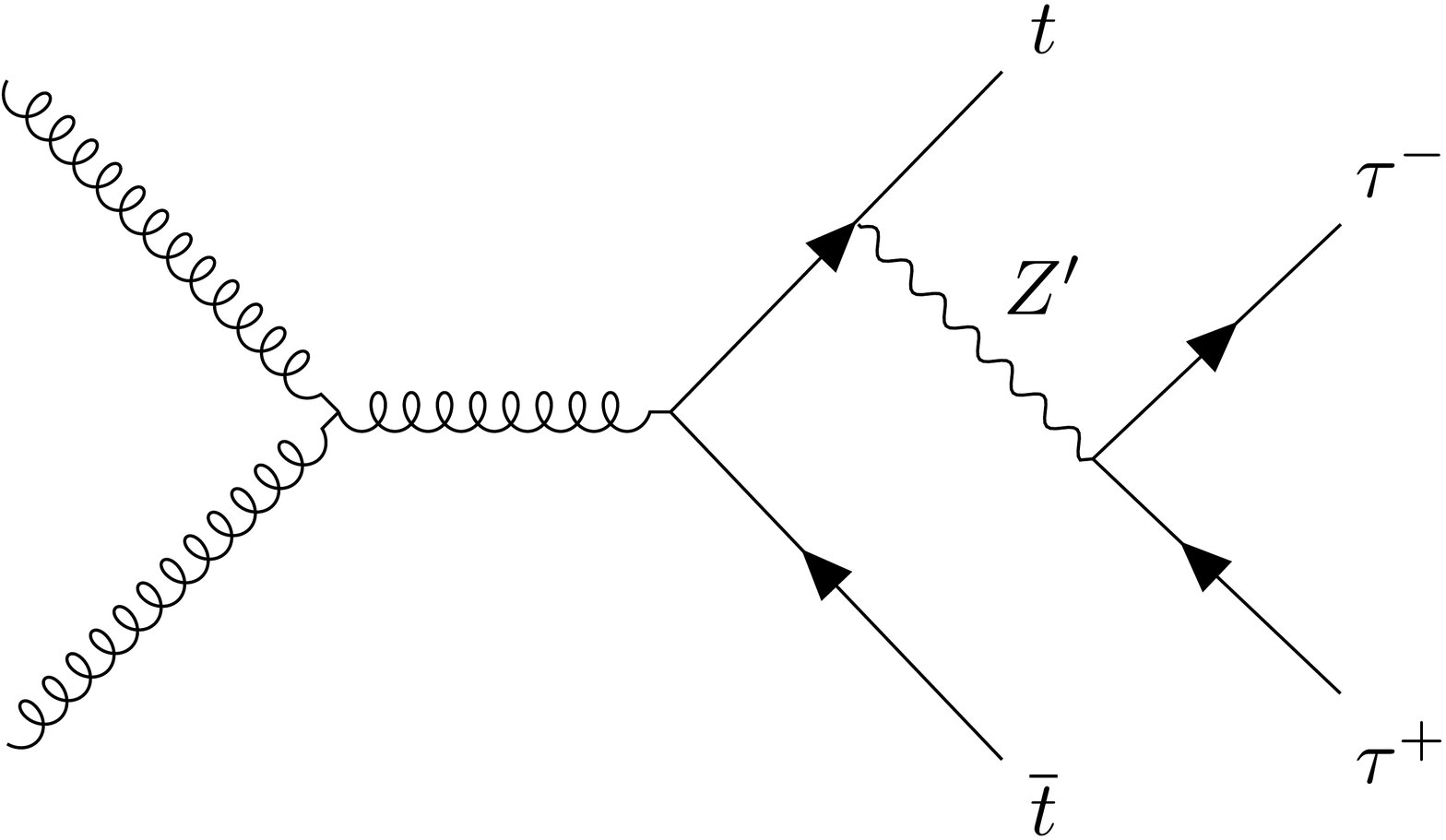}}
    \caption{\label{fig:feynman} Representative Feynman diagrams for $Z^\prime$ production via $t$-$\bar{t}$ fusion (left) and $t\bar{t}$ associated production (right).}
    \end{center}
\end{figure}

The $b\bar{b}Z^\prime$ production channel has been thoroughly investigated by Elahi \& Martin \cite{Elahi:2019} (from this point on referred to as EM19) using purely leptonic $\tau$ decay, and while it benefits from large statistics, it still suffers from these backgrounds as well as theoretical uncertainties in the proton's bottom-quark PDF. Production of $t\bar{t}Z^\prime$, driven by gluon-splitting into top quarks and gluon fusion (Fig.~\ref{fig:feynman}), trades large statistics for smaller PDF uncertainties and the opportunity to use single- or multi-lepton triggers on additional leptons from the $t \to W b$ decays. We take advantage of the higher lepton multiplicity available and consider semileptonic $W$ decays from $t\bar{t} \to b \bar{b} W^+W^-$ and semileptonic $\tau\tau$ decays, for $\tau_\ell \to e \nu_\tau \nu_e$ or $\tau_\ell \to \mu \nu_\tau \nu_\mu$; 

\begin{equation}
b\bar{b} W(\to \ell \nu_\ell) W(\to \bar{q} q )\tau_h \tau_\ell
\end{equation}

The three-lepton multiplicity of this selection is sufficient to almost entirely eliminate $t\bar{t}$ as a background, so we neglect it for the analysis. One could also consider $\tau_\ell \tau_\ell$ or $\tau_h \tau_h$, but we find that the former does not yield good fidelity in the $m_{Z^\prime}$ reconstruction and the latter does not sufficiently reduce the SM $t\bar{t}$ background. An important aspect of including a $\tau_h$ is that the $p_T$ threshold for $\tau_h$ triggers needed to sustain high efficiency selection at CMS and ATLAS ($p_T (\tau_h)\gtrsim 30$ GeV) is relatively higher than the triggers for $\ell = e,\mu$ \cite{CMS-DP-2019-020, CMS-DP-2018-044, CMS-DP-2018-006}. This is a major factor in the reduction of QCD and $t\bar{t}$ backgrounds, but effectively restricts us to $m_{Z^\prime} \gtrsim 30$ GeV. It is in this way that $b\bar{b}Z^\prime$ and $t\bar{t}Z^\prime$ searches may be complementary probes. 

The remaining irreducible backgrounds in this search come from $t\bar{t}Z$, $t\bar{t}W^\pm$, $t\bar{t}H$, and $t\bar{t}\gamma^*$ events. Since we consider light $m_{Z^\prime}$ near the $Z$ pole, the event content and relevant backgrounds may be very similar to that of a SM vector boson produced in association with top quarks. One can utilize some of the same strategies used in $t\bar{t}W$ or $t\bar{t}Z$ searches \cite{Aaboud:2019, Sirunyan:2018shy, Sirunyan:2018, CMS:2019nos}. From these analyses, we can motivate the choice of a three-lepton selection $\tau_h^\pm \ell^\mp \ell^\mp$ with two same-sign (SS) light leptons that are opposite in sign to the hadronically decaying tau lepton. The SS requirement serves to filter out opposite-sign (OS) signatures from SM $Z$ decays to light leptons. Also, for $m_{Z^\prime} \lesssim 50$ GeV, the $\tau_h$-$\tau_\ell$ system can be very proximate in $\Delta R = \sqrt{\Delta\eta^2 + \Delta\phi^2}$ (where $\eta$ is the pseudorapidity and $\phi$ is the transverse azimuthal angle of the detector geometry) and merge into a single $\tau_h$ object during reconstruction. In these cases, a leptonically decaying associated $W$, together with the merged $\tau_h$ object, could fake a semileptonic $\tau_h \tau_\ell$; the 3-lepton requirement greatly reduces this form of contamination. In order to choose the best light-lepton partner to the hadronic tau, we pick the one with the smallest $\Delta R(\tau_h^\pm, \ell^\mp)$ to disfavor isotropically distributed $\tau_h - W$ pairs. Additionally, although the pure $t\bar{t} + jets$ channel is partially suppressed by the 3 lepton requirement, the additional SS requirement on the light leptons ensures that the selection does not capture backgrounds arising from $t\bar{t} \rightarrow b \bar{b} \ell^+ \nu \ell^- \bar{\nu} + jets$ with one jet being misidentified as $\tau_h$.

On the hadronic side of the selection, we require at least three jets in the event and at least one of them $b$-tagged; since the $b$-tagging efficiency is $\sim60$\% ~\cite{Sirunyan:2017ezt} and the correct tag of the signal process is mostly dependent on the lepton selection and overall jet multiplicity, the single $b$ requirement helps increase acceptance without introducing sizable contamination.

Lastly, to ensure real sources of missing transverse energy ($\slashed{E}_T$) in the event from $W \to \ell \nu$ decays, we require $\slashed{E}_T > 30$ GeV. The kinematic cuts on $b$ jets, light jets, the $\tau_h$, the light ($e$/$\mu$) leptons, and the $\slashed{E}_T$ in each event are summarized in Table~\ref{table:selection}.  \\

We use \texttt{FeynRules} to generate a model file \cite{Christensen:2008py} \cite{Alloul:2013bka} and \texttt{MadGraph5 V2.6.4} \cite{Alwall:2014hca} to generate signal and background samples. \texttt{Pythia 8.2} \cite{Sjostrand:2014zea} is used for parton showering and \texttt{DELPHES 3.4} \cite{deFavereau:2013fsa} to model the detector response. We use a default \texttt{DELPHES} card modified to use a jet-finder cone radius of $\Delta R = 0.4$ and muon isolation cone of $\Delta R = 0.3$ restricting the summed $p_T$ of non-candidate tracks within the cone to less than 10\% of the candidate muon $p_T$.  For simplicity, we simulate signal and background at leading order (LO) and apply the ``$k$-factor" multipliers that account for the next-to-leading order (NLO) corrections for these processes \cite{Maltoni:2015ena}. Since our search strategy is designed for background suppression at the cost of few expected events, it is important to include these NLO corrections to offset any tree-level underestimation of the signal and background acceptances. These $k$-factors correspond to an increase of 22\% in the $t\bar{t}W^\pm$ cross-section and an overall increase of 26\% in the combined $t\bar{t}Z/H/\gamma^*$ cross-section. For the signal process, we approximate the $k$-factor as that of the $t\bar{t}Z$ process (+23\%).

\begin{center}
\begin{table}[]
    \centering
    \caption{Baseline selection criteria. $j_b$ is the number of selected $b$-jets.}
    \label{table:selection}
\begin{tabularx}{0.5\textwidth}{X X X l }
\hline \\
  & $|\eta| \leq$ & $p_T \geq$ & Multiplicity \\ \hline \\
 $\ell^\pm = e(\mu)$ & 2.1(2.4) &  26(23)  & 2 SS \\  
 $\tau_{h}^\mp $ & 2.4 &  30  & 1 \\
 $b$-jets & 2.4 &  20  & $\geq 1$ \\
 Light jets & 2.4 &  30  & $\geq$Max(3-$j_b$,0)\\
  $\slashed{E}_T$ & - & 30 & - \\\hline
 \end{tabularx}
\end{table}
\end{center}

\section{Results}
\label{sec:results}
\begin{figure}[!tbh]
  \begin{center}
   \includegraphics[width=0.5\textwidth]{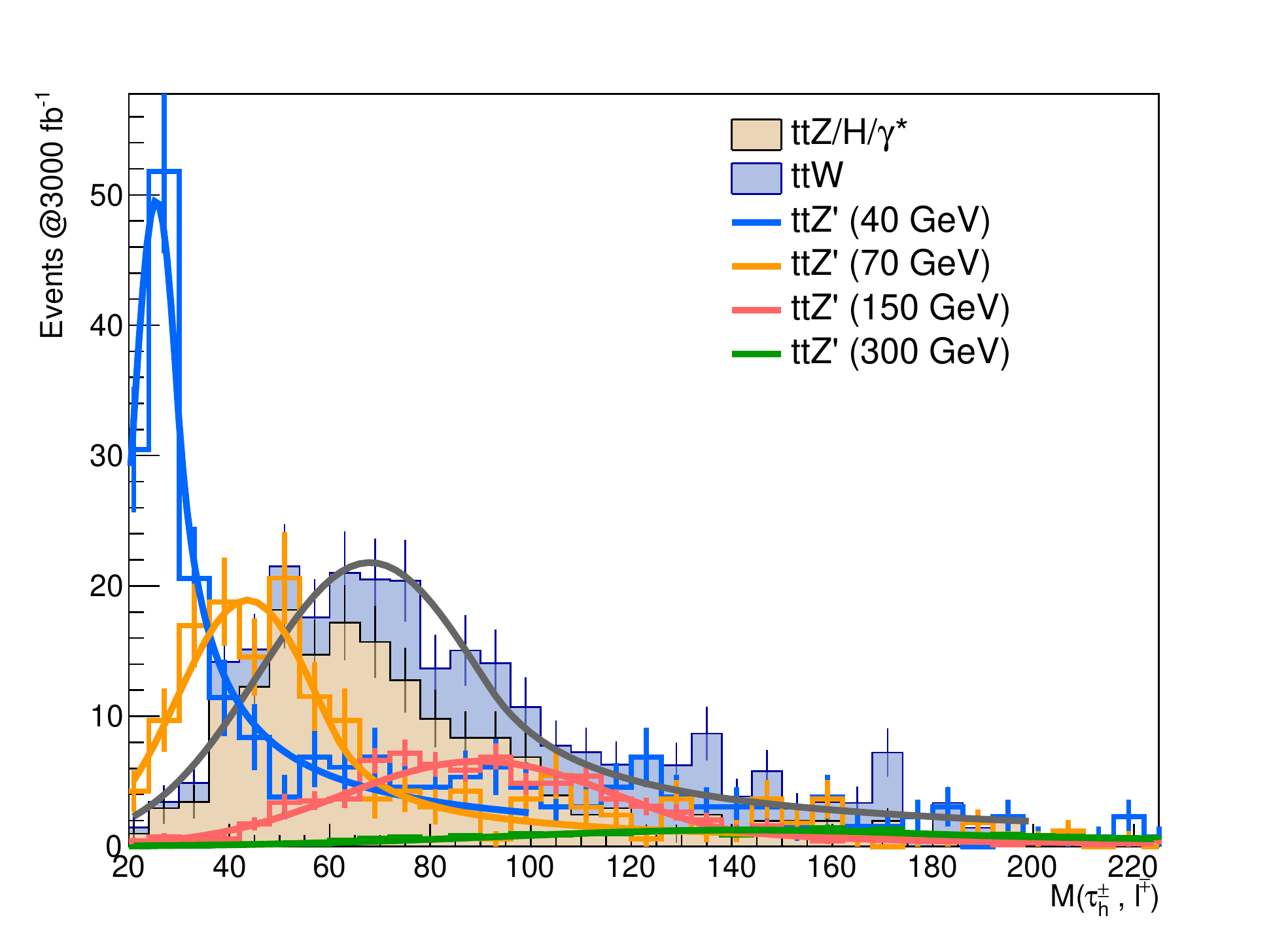}
    \caption{\label{fig:mass_spec} Mass spectra for $t\bar{t}Z^\prime$ in the semileptonic ditau channel with the counts and statistical error bars scaled to 3000 fb$^{-1}$ (before the $k$-factor NLO enhancement is applied). The signal samples were generated with couplings $g_t = 0.1$ and $g_\tau = 0.5$. Crystalball fits are shown as solid lines and the $t\bar{t}Z/H/\gamma^*$ and $t\bar{t}W$ backgrounds are stacked.}
    \end{center}
\end{figure}

The mass spectra generated after the selection procedure described in Section \ref{sec:search} is shown in Fig.~\ref{fig:mass_spec}. The usual invariant mass $m(\tau_h^\pm, \ell^\mp)$ is found to have the best shape determination and signal-to-background yield in comparison to various transverse-mass constructions such as $m_T (\tau_h^\pm, \ell^\mp)$ or $m_T(\tau_h^\pm, \ell^\mp, \slashed{E}_T)$. \\

\begin{figure}[tbh]
  \begin{center}
   \includegraphics[width=0.5\textwidth]{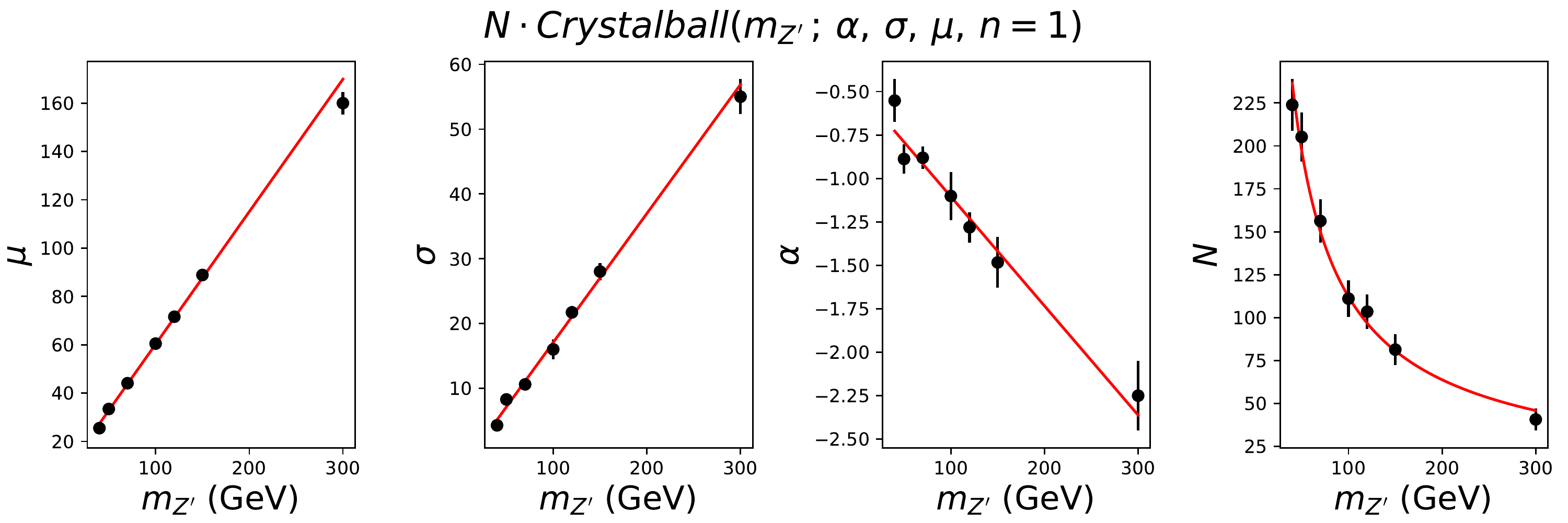}
    \caption{\label{fig:shapefits} Crystalball shape parameters from left to right: $\mu$ (mean), $\sigma$ (width), and $\alpha$ (gaussian-to-power-law transition parameter). The power law order parameter is fixed at $n=1$. On the far right we fit the integral over the invariant mass spectrum $N$ for couplings $g_q = 0.1$, $g_\tau = 0.5$ at 3000 fb$^{-1}$.}
    \end{center}
\end{figure}

To evaluate the projected sensitivity to this model at benchmark luminosities (300 and 3000 fb$^{-1}$), we construct a parametrically defined log-likelihood function of $g_q$ and $m_{Z^\prime}$, holding $g_\tau$ fixed without loss of generality. To do this, we first perform fits to the signal and background mass spectra and use them to model the asymptotic form of the mass spectrum for each point in model parameter space. \textit{Crystalball} distributions (constituting a power-law tail attached to a gaussian core) are used to fit the signal shape for seven samples $m_{Z^\prime}$ points between 40 and 300 GeV as well as the combined background shape; see Fig.~\ref{fig:mass_spec} as an example. Linear fits on the shape parameters as a function of mass allow us to interpolate smoothly between mass points. Additionally, to stabilize the interpolation, the power law order parameter $n$ of the Crystalball was fixed to natural numbers excluding zero, and for these choices the best set of fits was found at $n=1$. The normalization as a function of $m_{Z^\prime}$ was parameterized using a fit to $\propto m_{Z^\prime}^{-1}$. Changes in the coupling $g_t$ also shift the normalization up and down by a factor of $g_t^2$ following from the factor of $g_q^2 g_\tau^2$ in the $gg \to t\bar{t}Z^\prime (\tau^+ \tau^- )$ cross section. The Crystalball parameter fits are shown in Fig.~\ref{fig:shapefits}.\\

We define a log-likelihood likelihood function over the asymptotic binned events $N_{s+b} (g_q, m_{Z^\prime})$ tested against the background-only hypothesis $N_{b}\sim$ ($t\bar{t}W + t\bar{t}Z/\gamma^*/H$). We calculate the expected number of events by integrating the signal and background fits on $M(\tau_h^\pm, \ell^\mp)$ between bin edges. We test the signal-plus-background prediction, $\mu^i \equiv N^i_{s+b}(g_t, m_{Z^\prime})$, against the background prediction $\eta^i \equiv N^i_{b}$ in each bin $i$ for $g_\tau = 0.5$ fixed. The log-likelihood is given by the Poisson model in Equation~\ref{eq:loglike}. 
\begin{align}
\label{eq:loglike}
\mathcal{L} = \sum_{\text{bins}\,i} \bigg( \eta^i \ln \mu^i &- \mu^i - \ln \Gamma (\eta^i + 1)  \bigg)
\end{align}
Here we have used the Gamma function $\Gamma(\eta^i + 1)$ in order to analytically continue the usual factorial term to handle continuous numbers of events $\eta^i$. To perform the likelihood scan, we use the \texttt{MultiNest} bayesian inference package \cite{Feroz:2008xx} using flat priors on $g_t$ and $m_{Z^\prime}$ to determine the credible regions in $(m_{Z^\prime}, g_q)$ space. \texttt{MultiNest} evaluated the posterior distributions using an evidence tolerance of $0.1$ and a sample efficiency of $0.3$. 

\begin{figure}[tbh]
  \begin{center}
   \includegraphics[width=0.45\textwidth]{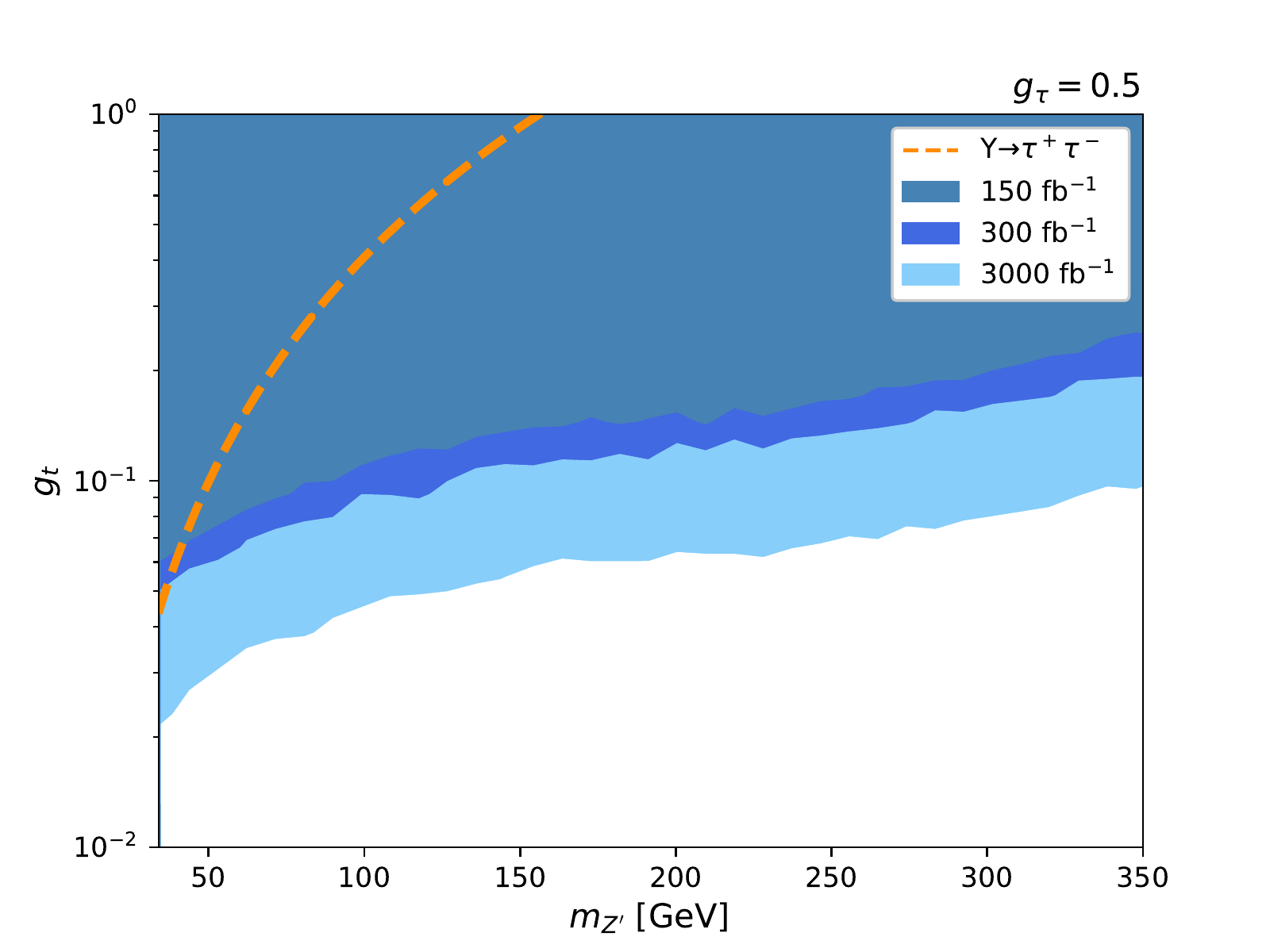}
    \caption{\label{fig:limits} Projected limits on $m_{Z^\prime}$ versus $g_t$ with fixed $g_\tau = 0.5$ for the minimal framework where the $Z^\prime$ boson couples only to right-handed top quarks and right handed tau leptons. The dark blue, blue, and light blue regions represent the projected 95\% exclusion from the LHC at $150$, $300$, and $3000$ fb$^{-1}$ of integrated luminosity, respectively. The limit from $R_{\mu\tau}$ in $\Upsilon$ decays (shown in dashed line) is valid only to models in which $g_t = g_b$.}
    \end{center}
\end{figure}

\begin{figure}[tbh]
  \begin{center}
   \includegraphics[width=0.45\textwidth]{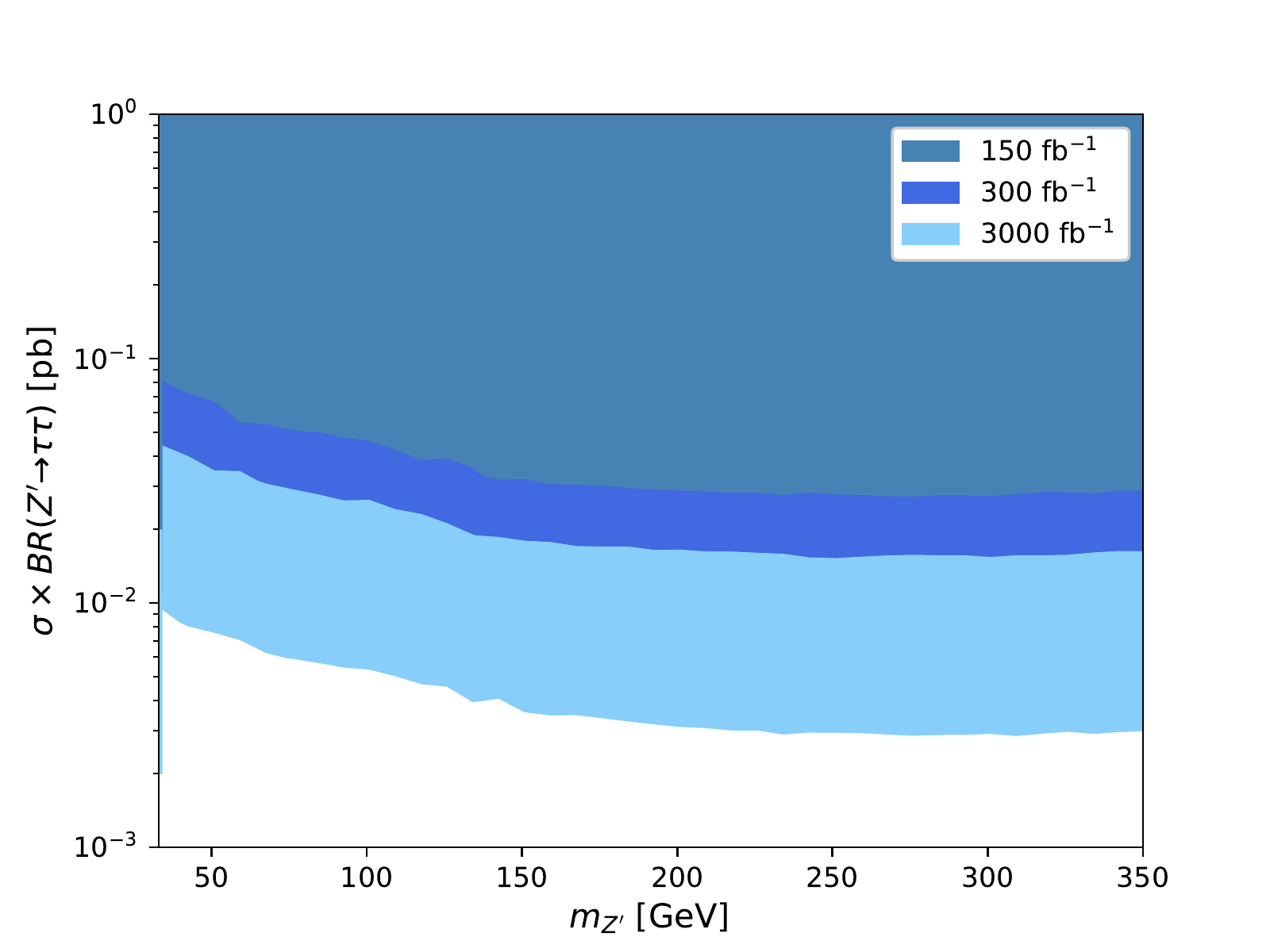}
    \caption{\label{fig:limits_xsbr} Projected limits on $m_{Z^\prime}$ versus the total cross-section times branching fraction $\sigma(pp \to Z^\prime) \times BR(Z^\prime \to \tau\tau)$ for the minimal framework where the $Z^\prime$ boson couples only to right-handed top quarks and right handed tau leptons. The dark blue, blue, and light blue regions represent the projected 95\% exclusion from the LHC at $150$, $300$, and $3000$ fb$^{-1}$ of integrated luminosity, respectively.}
    \end{center}
\end{figure}
The resulting 95\% ($\sim2\sigma$) credible exclusion on $g_t$ and $m_{Z^\prime}$ is shown in Fig.~\ref{fig:limits} for 150, 300, and 3000 fb$^{-1}$ of integrated luminosity. The tau coupling was fixed to $g_\tau = 0.5$, but one may easily extrapolate the calculated limits to a new value, $g_\tau^\prime$, by rescaling the exclusion curves as $g^\prime_t = (0.5 / g_\tau^\prime) g_t$.

The limit contrasts nicely with the EM19 analysis in that we are limited to $m_{Z^\prime} \gtrsim 30$ GeV but can achieve stable performance across masses greater than 80 GeV where EM19's analysis begins to lose performance. We remind the reader that how our results and EM19's fit together depends on the relationships between the various couplings in the full model. We also perform a scan over the cross-section times branching ratio $\sigma(pp \to Z^\prime) \times BR(Z^\prime \to \tau\tau)$, shown in Fig.~\ref{fig:limits_xsbr}. For this limit calculation we scale the signal shape according to $\bigg(\dfrac{\sigma \times BR(Z^\prime \to \tau\tau)}{\text{pb}}\bigg) \times \bigg(\dfrac{L}{\text{pb}^{-1}}\bigg) \times \epsilon(m)$ where $L$ is the luminosity and $\epsilon(m)$ is the signal efficiency parameterized by the $Z^\prime$ mass. The increased sensitivity with mass in Fig.~\ref{fig:limits_xsbr} is therefore purely due to the selection efficiency. 

Some existing search results may be re-interpreted as additional limits on the model considered here. For example, a search for pair-production of third-generation leptoquarks (LQ3’s) studies a final state with an electron or muon, two hadronically decaying tau leptons, and jets as a signature of the LQ3$\, \rightarrow t \tau$ decay \cite{Sirunyan:2018nkj}. The ``Category B" selection in this analysis began with requiring an isolated muon or electron with $p_T > 30$ GeV, an opposite-sign (OS) ditau pair with $p_T(\tau_h) > 65$ and 35 GeV, at least 3 jets, and missing transverse energy $\slashed{E}_T> 50$ GeV. We find the re-casted limits will not be as competitive as the limit derived from $\Upsilon$ decays at the BaBar experiment. Another example is a search for a pseudoscalar $A$ produced by $b\bar{b}$ fusion and decaying into tau leptons \cite{CMS:2019hvr}, which is complementary to our case, setting limits on the $g_b$ coupling of our $Z^\prime$ model.

\section{Conclusion}
We have investigated the phenomenology of the $Z^\prime$ boson decaying to tau leptons with exclusive third-generation couplings in the mass range $m_{Z^\prime} \in [30,350]$ GeV produced in association with top quarks in the high-luminosity era of the CERN LHC. Although our focus was placed on the $\tau_h \tau_\ell$ channel in order to suppress QCD and $t\bar{t}$ backgrounds using light lepton triggers, a study of how the $\tau_h \tau_h$ channel is worth pursuing. It should be noted that for right-handed top couplings, the $Z^\prime$ bosons are almost entirely unconstrained by existing experiments, opening the opportunity to make new exclusions of right-chiral models that exceed any existing constraints, using luminosities available by the early runs of the HL-LHC. The search strategy presented here should be complementary to one aiming to discover the $Z^\prime$ boson through $b\bar{b}$ production and decay to tau leptons.
\label{sec:conclusion}
\section*{Acknowledgements}
We thank B. Dutta for the motivating discussions and guidance in the model selection and search strategy, and we thank R. Kogler for the useful discussion on the di-$\tau_h$ channel. We also acknowledge the Texas A\&M University Brazos HPC cluster that contributed to the research reported here. 
M.A., T.K. and D.R. are supported in part by the Department of Energy grant de-sc0010813. M.D. and A.T. thank the Mitchell Institute for Fundamental Physics and Astronomy for support.
T.K. is also supported in part by the Qatar National Research Fund under project NPRP 9-328-1-066. 

\bibliographystyle{elsarticle-num}
\bibliography{main}

\begin{thebibliography}{10}
\expandafter\ifx\csname url\endcsname\relax
  \def\url#1{\texttt{#1}}\fi
\expandafter\ifx\csname urlprefix\endcsname\relax\def\urlprefix{URL }\fi
\expandafter\ifx\csname href\endcsname\relax
  \def\href#1#2{#2} \def\path#1{#1}\fi

\bibitem{Holdom:1994ss}
B.~Holdom, {Hints of new flavor physics at LEP?}, Phys. Lett. B. 339 (1994)
  114.
\newblock \href {http://arxiv.org/abs/hep-ph/9407311}
  {\path{arXiv:hep-ph/9407311}}, \href
  {http://dx.doi.org/10.1016/0370-2693(94)91142-8}
  {\path{doi:10.1016/0370-2693(94)91142-8}}.

\bibitem{Frampton:1996bs}
P.~H. Frampton, M.~B. Wise, B.~D. Wright, {An Elusive Z-prime coupled to
  beauty}, Phys. Rev. D. 54 (1996) 5820--5823.
\newblock \href {http://arxiv.org/abs/hep-ph/9604260}
  {\path{arXiv:hep-ph/9604260}}, \href
  {http://dx.doi.org/10.1103/PhysRevD.54.5820}
  {\path{doi:10.1103/PhysRevD.54.5820}}.

\bibitem{Muller:1996dj}
D.~J. Muller, S.~Nandi, {Top flavor: A Separate SU(2) for the third family},
  Phys. Lett. B. 383 (1996) 345--350.
\newblock \href {http://arxiv.org/abs/hep-ph/9602390}
  {\path{arXiv:hep-ph/9602390}}, \href
  {http://dx.doi.org/10.1016/0370-2693(96)00745-9}
  {\path{doi:10.1016/0370-2693(96)00745-9}}.

\bibitem{Andrianov:1998hx}
A.~A. Andrianov, P.~Osland, A.~A. Pankov, N.~V. Romanenko, J.~Sirkka, {On the
  phenomenology of a $Z^\prime$ coupling only to third family fermions}, Phys.
  Rev. D. 58 (1998) 075001.
\newblock \href {http://arxiv.org/abs/hep-ph/9804389}
  {\path{arXiv:hep-ph/9804389}}, \href
  {http://dx.doi.org/10.1103/PhysRevD.58.075001}
  {\path{doi:10.1103/PhysRevD.58.075001}}.

\bibitem{Babu:2017olk}
K.~S. Babu, A.~Friedland, P.~A.~N. Machado, I.~Mocioiu, {Flavor Gauge Models
  Below the Fermi Scale}, JHEP 12 (2017) 096.
\newblock \href {http://arxiv.org/abs/1705.01822} {\path{arXiv:1705.01822}},
  \href {http://dx.doi.org/10.1007/JHEP12(2017)096}
  {\path{doi:10.1007/JHEP12(2017)096}}.

\bibitem{Kamada:2018kmi}
A.~Kamada, M.~Yamada, T.~T. Yanagida, {Self-interacting dark matter with a
  vector mediator: kinetic mixing with the $
  \mathrm{U}{(1)}_{{\left(B-L\right)}_3} $ gauge boson}, JHEP 03 (2019) 021.
\newblock \href {http://arxiv.org/abs/1811.02567} {\path{arXiv:1811.02567}},
  \href {http://dx.doi.org/10.1007/JHEP03(2019)021}
  {\path{doi:10.1007/JHEP03(2019)021}}.

\bibitem{Elahi:2019}
F.~Elahi, A.~Martin, {LHC constraints on a $(B-L)_3$ gauge boson}, Phys. Rev.
  D. 100 (2019) 035016.
\newblock \href {http://arxiv.org/abs/1905.10106} {\path{arXiv:1905.10106}},
  \href {http://dx.doi.org/10.1103/PhysRevD.100.035016}
  {\path{doi:10.1103/PhysRevD.100.035016}}.

\bibitem{Aad:2015osa}
G.~Aad, et~al., {A search for high-mass resonances decaying to
  $\tau^{+}\tau^{-}$ in $pp$ collisions at $\sqrt{s}=8$ TeV with the ATLAS
  detector}, JHEP 07 (2015) 157.
\newblock \href {http://arxiv.org/abs/1502.07177} {\path{arXiv:1502.07177}},
  \href {http://dx.doi.org/10.1007/JHEP07(2015)157}
  {\path{doi:10.1007/JHEP07(2015)157}}.

\bibitem{CMS:2016zxk}
V.~Khachatryan, et~al., {Search for heavy resonances decaying to tau lepton
  pairs in proton-proton collisions at $ \sqrt{s}=13 $ TeV}, JHEP 02 (2017)
  048.
\newblock \href {http://arxiv.org/abs/1611.06594} {\path{arXiv:1611.06594}},
  \href {http://dx.doi.org/10.1007/JHEP02(2017)048}
  {\path{doi:10.1007/JHEP02(2017)048}}.

\bibitem{Sirunyan:2017uhk}
A.~M. Sirunyan, et~al., {Search for $ \mathrm{t}\overline{\mathrm{t}} $
  resonances in highly boosted lepton+jets and fully hadronic final states in
  proton-proton collisions at $ \sqrt{s}=13 $ TeV}, JHEP 07 (2017) 001.
\newblock \href {http://arxiv.org/abs/1704.03366} {\path{arXiv:1704.03366}},
  \href {http://dx.doi.org/10.1007/JHEP07(2017)001}
  {\path{doi:10.1007/JHEP07(2017)001}}.

\bibitem{Haller:2017hjf}
J.~Haller, et~al., {Searches for new resonances with couplings to third
  generation quarks with the CMS detector}, PoS EPS-HEP2017 (2017) 289.
\newblock \href {http://dx.doi.org/10.22323/1.314.0289}
  {\path{doi:10.22323/1.314.0289}}.

\bibitem{ATLAS:2017mpg}
M.~Aaboud, et~al., \href{http://dx.doi.org/10.1007/JHEP01(2018)055}{Search for
  additional heavy neutral higgs and gauge bosons in the ditau final state
  produced in 36 fb$^{-1}$ of $pp$ collisions at $ \sqrt{s}=13 $ tev with the
  atlas detector}, JHEP 01 (2018) 055.
\newblock \href {http://arxiv.org/abs/1709.07242} {\path{arXiv:1709.07242}},
  \href {http://dx.doi.org/10.1007/jhep01(2018)055}
  {\path{doi:10.1007/jhep01(2018)055}}.
\newline\urlprefix\url{http://dx.doi.org/10.1007/JHEP01(2018)055}

\bibitem{Sirunyan:2018ikr}
A.~M. Sirunyan, et~al., {Search for low-mass resonances decaying into bottom
  quark-antiquark pairs in proton-proton collisions at $\sqrt{s} =$ 13 TeV},
  Phys. Rev. D. 99 (2019) 012005.
\newblock \href {http://arxiv.org/abs/1810.11822} {\path{arXiv:1810.11822}},
  \href {http://dx.doi.org/10.1103/PhysRevD.99.012005}
  {\path{doi:10.1103/PhysRevD.99.012005}}.

\bibitem{CMS:2019hvr}
A.~M. Sirunyan, et~al., {Search for a low-mass $\tau^+\tau^-$ resonance in
  association with a bottom quark in proton-proton collisions at $\sqrt{s}=$ 13
  TeV}, JHEP 05 (2019) 210.
\newblock \href {http://arxiv.org/abs/1903.10228} {\path{arXiv:1903.10228}},
  \href {http://dx.doi.org/10.1007/JHEP05(2019)210}
  {\path{doi:10.1007/JHEP05(2019)210}}.

\bibitem{Hill:1991at}
C.~T. Hill, {Topcolor: Top quark condensation in a gauge extension of the
  standard model}, Phys. Lett. B. 266 (1991) 419.
\newblock \href {http://dx.doi.org/10.1016/0370-2693(91)91061-Y}
  {\path{doi:10.1016/0370-2693(91)91061-Y}}.

\bibitem{Hill:1993hs}
C.~T. Hill, S.~J. Parke, {Top production: Sensitivity to new physics}, Phys.
  Rev. D. 49 (1994) 4454--4462.
\newblock \href {http://arxiv.org/abs/hep-ph/9312324}
  {\path{arXiv:hep-ph/9312324}}, \href
  {http://dx.doi.org/10.1103/PhysRevD.49.4454}
  {\path{doi:10.1103/PhysRevD.49.4454}}.

\bibitem{Hill:1994hp}
C.~T. Hill, {Topcolor assisted technicolor}, Phys. Lett. B. 345 (1995) 483.
\newblock \href {http://arxiv.org/abs/hep-ph/9411426}
  {\path{arXiv:hep-ph/9411426}}, \href
  {http://dx.doi.org/10.1016/0370-2693(94)01660-5}
  {\path{doi:10.1016/0370-2693(94)01660-5}}.

\bibitem{Harris:2011ez}
R.~M. Harris, S.~Jain, {Cross Sections for Leptophobic Topcolor Z' Decaying to
  Top-Antitop}, Eur. Phys. J. C. 72 (2012) 2072.
\newblock \href {http://arxiv.org/abs/1112.4928} {\path{arXiv:1112.4928}},
  \href {http://dx.doi.org/10.1140/epjc/s10052-012-2072-4}
  {\path{doi:10.1140/epjc/s10052-012-2072-4}}.

\bibitem{Rosner:1996eb}
J.~L. Rosner, {Prominent decay modes of a leptophobic $Z^\prime$}, Phys. Lett.
  B. 387 (1996) 113.
\newblock \href {http://arxiv.org/abs/hep-ph/9607207}
  {\path{arXiv:hep-ph/9607207}}, \href
  {http://dx.doi.org/10.1016/0370-2693(96)01022-2}
  {\path{doi:10.1016/0370-2693(96)01022-2}}.

\bibitem{Lynch:2000md}
K.~R. Lynch, E.~H. Simmons, M.~Narain, S.~Mrenna, {Finding $Z^\prime$ bosons
  coupled preferentially to the third family at LEP and the Tevatron}, Phys.
  Rev. D. 63 (2001) 035006.
\newblock \href {http://arxiv.org/abs/hep-ph/0007286}
  {\path{arXiv:hep-ph/0007286}}, \href
  {http://dx.doi.org/10.1103/PhysRevD.63.035006}
  {\path{doi:10.1103/PhysRevD.63.035006}}.

\bibitem{Carena:2004xs}
M.~Carena, A.~Daleo, B.~A. Dobrescu, T.~M.~P. Tait, {$Z^\prime$ gauge bosons at
  the Tevatron}, Phys. Rev. D. 70 (2004) 093009.
\newblock \href {http://arxiv.org/abs/hep-ph/0408098}
  {\path{arXiv:hep-ph/0408098}}, \href
  {http://dx.doi.org/10.1103/PhysRevD.70.093009}
  {\path{doi:10.1103/PhysRevD.70.093009}}.

\bibitem{Choudhury:2007ux}
D.~Choudhury, R.~M. Godbole, R.~K. Singh, K.~Wagh, {Top production at the
  Tevatron/LHC and nonstandard, strongly interacting spin one particles}, Phys.
  Lett. B. 657 (2007) 69.
\newblock \href {http://arxiv.org/abs/0705.1499} {\path{arXiv:0705.1499}},
  \href {http://dx.doi.org/10.1016/j.physletb.2007.09.057}
  {\path{doi:10.1016/j.physletb.2007.09.057}}.

\bibitem{Khachatryan:2015sma}
V.~Khachatryan, et~al., {Search for resonant $t \bar t$ production in
  proton-proton collisions at $\sqrt s=$ 8 TeV}, Phys. Rev. D. 93 (2016)
  012001.
\newblock \href {http://arxiv.org/abs/1506.03062} {\path{arXiv:1506.03062}},
  \href {http://dx.doi.org/10.1103/PhysRevD.93.012001}
  {\path{doi:10.1103/PhysRevD.93.012001}}.

\bibitem{CMS:2018ohu}
\href{http://cds.cern.ch/record/2649032}{{Search for $t\bar{t}$ resonances at
  the HL-LHC and HE-LHC with the Phase-2 CMS detector}}, Tech. Rep.
  CMS-PAS-FTR-18-009, CERN, Geneva (2018).
\newline\urlprefix\url{http://cds.cern.ch/record/2649032}

\bibitem{Aaboud:2018mjh}
M.~Aaboud, et~al., {Search for heavy particles decaying into top-quark pairs
  using lepton-plus-jets events in proton-proton collisions at $\sqrt{s} = 13$
  $\text {TeV}$ with the ATLAS detector}, Eur. Phys. J. C. 78 (2018) 565.
\newblock \href {http://arxiv.org/abs/1804.10823} {\path{arXiv:1804.10823}},
  \href {http://dx.doi.org/10.1140/epjc/s10052-018-5995-6}
  {\path{doi:10.1140/epjc/s10052-018-5995-6}}.

\bibitem{Sirunyan:2017yar}
A.~M. Sirunyan, et~al., {Measurement of the jet mass in highly boosted
  ${\mathrm{t}}\overline{\mathrm{t}}$ events from pp collisions at $\sqrt{s}=8$
  $\,\text {TeV}$}, Eur. Phys. J. C. 77 (2017) 467.
\newblock \href {http://arxiv.org/abs/1703.06330} {\path{arXiv:1703.06330}},
  \href {http://dx.doi.org/10.1140/epjc/s10052-017-5030-3}
  {\path{doi:10.1140/epjc/s10052-017-5030-3}}.

\bibitem{Cerrito:2016qig}
L.~Cerrito, et~al., {Discovering and profiling $Z^\prime$ bosons using
  asymmetry observables in top quark pair production with the lepton-plus-jets
  final state at the LHC}\href {http://arxiv.org/abs/1609.05540}
  {\path{arXiv:1609.05540}}.

\bibitem{Arina:2016cqj}
C.~Arina, et~al., {A comprehensive approach to dark matter studies: exploration
  of simplified top-philic models}, JHEP 11 (2016) 111.
\newblock \href {http://arxiv.org/abs/1605.09242} {\path{arXiv:1605.09242}},
  \href {http://dx.doi.org/10.1007/JHEP11(2016)111}
  {\path{doi:10.1007/JHEP11(2016)111}}.

\bibitem{Pedersen:2015knf}
K.~Pedersen, Z.~Sullivan, {$\mu_x$ boosted-bottom-jet tagging and Z′ boson
  searches}, Phys. Rev. D. 93 (2016) 014014.
\newblock \href {http://arxiv.org/abs/1511.05990} {\path{arXiv:1511.05990}},
  \href {http://dx.doi.org/10.1103/PhysRevD.93.014014}
  {\path{doi:10.1103/PhysRevD.93.014014}}.

\bibitem{Fox:2018ldq}
P.~J. Fox, I.~Low, Y.~Zhang, {Top-philic $Z'$ forces at the LHC}, JHEP 03
  (2018) 074.
\newblock \href {http://arxiv.org/abs/1801.03505} {\path{arXiv:1801.03505}},
  \href {http://dx.doi.org/10.1007/JHEP03(2018)074}
  {\path{doi:10.1007/JHEP03(2018)074}}.

\bibitem{Alvarez:2019uxp}
E.~Alvarez, R.~M.~S. Seoane, A.~Juste, {Four-top as probe of light top-philic
  New Physics}\href {http://arxiv.org/abs/1910.09581}
  {\path{arXiv:1910.09581}}.

\bibitem{Boucenna:2016qad}
S.~M. Boucenna, A.~Celis, J.~Fuentes-Martin, A.~Vicente, J.~Virto,
  {Phenomenology of an $SU(2) \times SU(2) \times U(1)$ model with
  lepton-flavour non-universality}, JHEP 12 (2016) 059.
\newblock \href {http://arxiv.org/abs/1608.01349} {\path{arXiv:1608.01349}},
  \href {http://dx.doi.org/10.1007/JHEP12(2016)059}
  {\path{doi:10.1007/JHEP12(2016)059}}.

\bibitem{PhysRevLett.104.191801}
del Amo~Sanchez, et~al.,
  \href{https://link.aps.org/doi/10.1103/PhysRevLett.104.191801}{{Test of
  Lepton Universality in $\ensuremath{\Upsilon}(1S)$ Decays at BABAR}}, Phys.
  Rev. Lett. 104 (2010) 191801.
\newblock \href {http://dx.doi.org/10.1103/PhysRevLett.104.191801}
  {\path{doi:10.1103/PhysRevLett.104.191801}}.
\newline\urlprefix\url{https://link.aps.org/doi/10.1103/PhysRevLett.104.191801}

\bibitem{Aloni:2017eny}
D.~Aloni, A.~Efrati, Y.~Grossman, Y.~Nir, {$\Upsilon$ and $\psi$ leptonic
  decays as probes of solutions to the $R_D^{(*)}$ puzzle}, JHEP 06 (2017) 019.
\newblock \href {http://arxiv.org/abs/1702.07356} {\path{arXiv:1702.07356}},
  \href {http://dx.doi.org/10.1007/JHEP06(2017)019}
  {\path{doi:10.1007/JHEP06(2017)019}}.

\bibitem{Sirunyan:2018pgf}
A.~M. Sirunyan, et~al., {Performance of reconstruction and identification of
  $\tau$ leptons decaying to hadrons and $\nu_\tau$ in pp collisions at
  $\sqrt{s}=$ 13 TeV}, JINST 13 (2018) P10005.
\newblock \href {http://arxiv.org/abs/1809.02816} {\path{arXiv:1809.02816}},
  \href {http://dx.doi.org/10.1088/1748-0221/13/10/P10005}
  {\path{doi:10.1088/1748-0221/13/10/P10005}}.

\bibitem{CMS-DP-2019-020}
\href{https://cds.cern.ch/record/2682202}{{Level-1 E/Gamma trigger performance
  in Run II}}.
\newline\urlprefix\url{https://cds.cern.ch/record/2682202}

\bibitem{CMS-DP-2018-044}
\href{https://cds.cern.ch/record/2629852}{{Level-1 Muon Trigger Performance}}.
\newline\urlprefix\url{https://cds.cern.ch/record/2629852}

\bibitem{CMS-DP-2018-006}
\href{https://cds.cern.ch/record/2305547}{{Level-1 $\tau$ trigger performance
  in 2017 data}}.
\newline\urlprefix\url{https://cds.cern.ch/record/2305547}

\bibitem{Aaboud:2019}
M.~Aaboud, et~al.,
  \href{https://link.aps.org/doi/10.1103/PhysRevD.99.072009}{{Measurement of
  the $t\overline{t}Z$ and $t\overline{t}W$ cross sections in proton-proton
  collisions at $\sqrt{s}=13\text{ }\text{ }\mathrm{TeV}$ with the ATLAS
  detector}}, Phys. Rev. D 99 (2019) 072009.
\newblock \href {http://dx.doi.org/10.1103/PhysRevD.99.072009}
  {\path{doi:10.1103/PhysRevD.99.072009}}.
\newline\urlprefix\url{https://link.aps.org/doi/10.1103/PhysRevD.99.072009}

\bibitem{Sirunyan:2018shy}
A.~M. Sirunyan, et~al., {Evidence for associated production of a Higgs boson
  with a top quark pair in final states with electrons, muons, and hadronically
  decaying $\tau$ leptons at $\sqrt{s} =$ 13 TeV}, JHEP 08 (2018) 066.
\newblock \href {http://arxiv.org/abs/1803.05485} {\path{arXiv:1803.05485}},
  \href {http://dx.doi.org/10.1007/JHEP08(2018)066}
  {\path{doi:10.1007/JHEP08(2018)066}}.

\bibitem{Sirunyan:2018}
A.~Sirunyan, et~al.,
  \href{https://doi.org/10.1007/JHEP08(2018)011}{{Measurement of the cross
  section for top quark pair production in association with a W or Z boson in
  proton-proton collisions at $\sqrt{s}=13$ TeV}}, JHEP 08 (2018) 11.
\newblock \href {http://arxiv.org/abs/1711.02547} {\path{arXiv:1711.02547}},
  \href {http://dx.doi.org/10.1007/JHEP08(2018)011}
  {\path{doi:10.1007/JHEP08(2018)011}}.
\newline\urlprefix\url{https://doi.org/10.1007/JHEP08(2018)011}

\bibitem{CMS:2019nos}
A.~M. Sirunyan, et~al., {Measurement of top quark pair production in
  association with a Z boson in proton-proton collisions at $\sqrt{s}=$ 13
  TeV}\href {http://arxiv.org/abs/1907.11270} {\path{arXiv:1907.11270}}.

\bibitem{Sirunyan:2017ezt}
A.~M. Sirunyan, et~al., {Identification of heavy-flavour jets with the CMS
  detector in pp collisions at 13 TeV}, JINST 13 (2018) P05011.
\newblock \href {http://arxiv.org/abs/1712.07158} {\path{arXiv:1712.07158}},
  \href {http://dx.doi.org/10.1088/1748-0221/13/05/P05011}
  {\path{doi:10.1088/1748-0221/13/05/P05011}}.

\bibitem{Christensen:2008py}
N.~D. Christensen, C.~Duhr, {FeynRules - Feynman rules made easy}, Comput.
  Phys. Commun. 180 (2009) 1614--1641.
\newblock \href {http://arxiv.org/abs/0806.4194} {\path{arXiv:0806.4194}},
  \href {http://dx.doi.org/10.1016/j.cpc.2009.02.018}
  {\path{doi:10.1016/j.cpc.2009.02.018}}.

\bibitem{Alloul:2013bka}
A.~Alloul, N.~D. Christensen, C.~Degrande, C.~Duhr, B.~Fuks, {FeynRules 2.0 - A
  complete toolbox for tree-level phenomenology}, Comput. Phys. Commun. 185
  (2014) 2250.
\newblock \href {http://arxiv.org/abs/1310.1921} {\path{arXiv:1310.1921}},
  \href {http://dx.doi.org/10.1016/j.cpc.2014.04.012}
  {\path{doi:10.1016/j.cpc.2014.04.012}}.

\bibitem{Alwall:2014hca}
J.~Alwall, et~al., {The automated computation of tree-level and next-to-leading
  order differential cross sections, and their matching to parton shower
  simulations}, JHEP 07 (2014) 079.
\newblock \href {http://arxiv.org/abs/1405.0301} {\path{arXiv:1405.0301}},
  \href {http://dx.doi.org/10.1007/JHEP07(2014)079}
  {\path{doi:10.1007/JHEP07(2014)079}}.

\bibitem{Sjostrand:2014zea}
T.~Sjostrand, S.~Ask, J.~R. Christiansen, R.~Corke, N.~Desai, P.~Ilten,
  S.~Mrenna, S.~Prestel, C.~O. Rasmussen, P.~Z. Skands, {An Introduction to
  PYTHIA 8.2}, Comput. Phys. Commun. 191 (2015) 159.
\newblock \href {http://arxiv.org/abs/1410.3012} {\path{arXiv:1410.3012}},
  \href {http://dx.doi.org/10.1016/j.cpc.2015.01.024}
  {\path{doi:10.1016/j.cpc.2015.01.024}}.

\bibitem{deFavereau:2013fsa}
J.~de~Favereau, C.~Delaere, P.~Demin, A.~Giammanco, V.~Lema\`{i}tre,
  A.~Mertens, M.~Selvaggi, {DELPHES 3, A modular framework for fast simulation
  of a generic collider experiment}, JHEP 02 (2014) 057.
\newblock \href {http://arxiv.org/abs/1307.6346} {\path{arXiv:1307.6346}},
  \href {http://dx.doi.org/10.1007/JHEP02(2014)057}
  {\path{doi:10.1007/JHEP02(2014)057}}.

\bibitem{Maltoni:2015ena}
F.~Maltoni, D.~Pagani, I.~Tsinikos, {Associated production of a top-quark pair
  with vector bosons at NLO in QCD: impact on $t\bar{t}H$ searches at the LHC},
  JHEP 02 (2016) 113.
\newblock \href {http://arxiv.org/abs/1507.05640} {\path{arXiv:1507.05640}},
  \href {http://dx.doi.org/10.1007/JHEP02(2016)113}
  {\path{doi:10.1007/JHEP02(2016)113}}.

\bibitem{Feroz:2008xx}
F.~Feroz, M.~P. Hobson, M.~Bridges, {MultiNest: an efficient and robust
  Bayesian inference tool for cosmology and particle physics}, Mon. Not. Roy.
  Astron. Soc. 398 (2009) 1601.
\newblock \href {http://arxiv.org/abs/0809.3437} {\path{arXiv:0809.3437}},
  \href {http://dx.doi.org/10.1111/j.1365-2966.2009.14548.x}
  {\path{doi:10.1111/j.1365-2966.2009.14548.x}}.

\bibitem{Sirunyan:2018nkj}
A.~M. Sirunyan, et~al., {Search for third-generation scalar leptoquarks
  decaying to a top quark and a $\tau$ lepton at $\sqrt{s}=$ 13 TeV}, Eur.
  Phys. J. C. 78 (2018) 707.
\newblock \href {http://arxiv.org/abs/1803.02864} {\path{arXiv:1803.02864}},
  \href {http://dx.doi.org/10.1140/epjc/s10052-018-6143-z}
  {\path{doi:10.1140/epjc/s10052-018-6143-z}}.

\end{thebibliography}

\end{document}